# Film Thickness Gauge Based on Interferometric Principle of Y-shaped Optical Fiber

Juntao He[1#], Yikai Dang[1#], Haoqi Wang[1#], Shaohua Wang[1#], Yingke Li[1#], Ruiyun Ma[1],

Yingyuan Li[1], Peilin Gao[1], Jianguo Cao[1], Yong Pan[1]*

1-College of Science, Xi'an University of Architecture and Technology, Xi'an 710055, China

*Corresponding author: panyong@xauat.edu.cn

**Abstract:** In this paper, a thin film thickness gauge based on the interferometric principle of Y-shaped optical fiber is proposed to achieve accurate measurement of film thickness. In this paper, the optical fiber, the interferometric principle and the film thickness calculation principle are introduced, and the interferometric thickness measurement system based on Y-shaped optical fiber is constructed. The system uses the special structure of Y-shaped optical fiber to transmit the optical signal generated by the light source to the surface of the thin film, and obtains coherent optical signals of different wavelengths through reflection and interference. The spectrometer is used to receive and interpret these interference signals, and the thickness of the film is calculated according to the wavelength difference of the peak positions of the adjacent stages, combined with the refractive index of the film. In the specific design, the paper elaborates on the design of each part of the instrument, including the selection and parameter setting of the light source, Y-fiber and spectrometer. Among them, the Y-shaped optical fiber, as the core component of the instrument, has the function of transmitting optical signals and detecting optical signals on the surface of thin films. At the same time, the paper also introduces the housing packaging and internal assembly process of the instrument to ensure the portability and stability of the instrument. The results show that the thickness gauge has high measurement accuracy and stability, which can meet the needs of practical applications.

**Keywords:** Film thickness gauge; Y-shaped optical fiber; Interferometric principle; Grating spectrometer

# 1.Introduction

In recent years, photonic chips, quantum computing and other emerging fields of information technology have attracted much attention, and [1-3] have been highly valued by various countries. At present, it is still necessary to achieve a breakthrough to obtain the low-dimensional light source matching the current light/quantum chip[4-6]. The key problem is that the measurement of the dimensions and thickness of various materials is not accurate, convenient, and the cost is too high [7-10]. As an important way to improve the performance of devices, thin film is widely used in modern optical, electronic, medical and other technical fields. With the progress of thin film preparation process, the large area modular production of nano thin film has become mature, and nano thin thickness, complex dispersion relationship of measurement technology and method put forward the great challenge [11], the optical thin film thickness measurement method is still unable to balance high precision, light volume, and reasonable system cost of [12].

White light interference technology is a common optical measurement method, widely used in film thickness measurement, surface morphology measurement, refractive index measurement and other fields. The white light fiber interferometric membrane thickness instrument is a special instrument designed based on the principle of white light interference technology to measure the thickness of the membrane layer. At present, the mainstream membrane thickness instrument methods used in the market are as follows:

Elliptic polarization method[13-16]: This method determines the thickness and refractive index of the film by measuring the reflection and transmission characteristics of the elliptically polarized light. Using the phenomenon of incident reflection and interference on the film and substrate surface, the thickness and the optical constant of the film can be calculated by analyzing the amplitude and phase changes of the elliptically polarized light. Interferometry[17-19]: Use the interference phenomenon of light to measure the thickness of the film, usually by observing the interference stripes of light or by using an interferometer. The film can cause the interference phenomenon of incident light, and the film thickness can be calculated by measuring the spacing of the interference fringes or the output signal of the interferometer.

## 2.Detection principle

Interference phenomenon is a unique feature of fluctuation phenomenon. When two or multiple waves meet in space and overlap on each other, the phenomenon of vibration is always strengthened in some regions, while the vibration is always weakened in other regions, as shown in Figure 1 A. The interference of light is one of the interference phenomena. Thin-film interference is a manifestation of the interference phenomenon of light on the thin-film medium. When a light wave hits the film, it is reflected at the two interfaces due to the different refractive index at the upper and lower interfaces of the film. These reflected light waves will interfere with each other to form a new light wave.

An incident beam is illuminated vertically radiated on the film to be measured and the optical path diagram is shown in Figure 1 B.

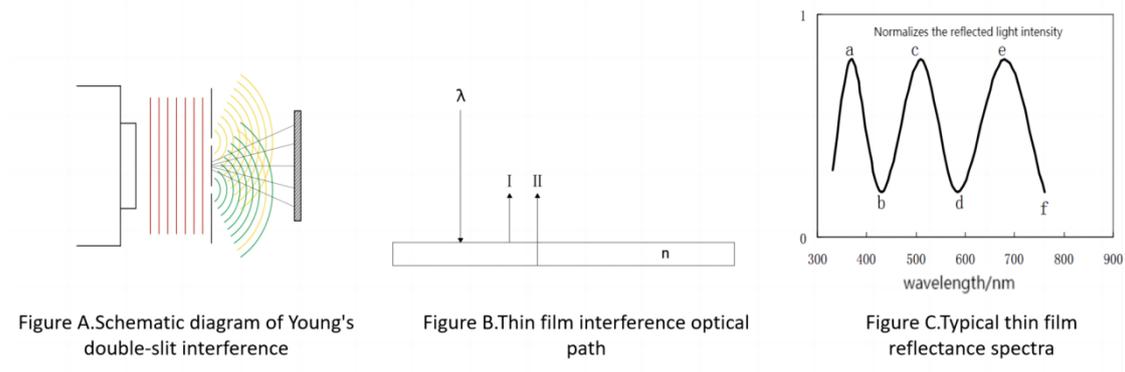

Figure A.Schematic diagram of Young's double-slit interference

Figure B.Thin film interference optical path

Figure C.Typical thin film reflectance spectra

Figure 1

Because the thickness of the film to be measured is uniform, the difference between the two beams in the film is equal

$$\Delta L = 2nd \qquad (1-1)$$

If the phase difference $\Delta\varphi(\Delta\varphi=\varphi_2-\varphi_1)$is constant for each other at each specified point in the light field, the resultant light intensity at a point in the encounter space is

$$I = I_1 + I_2 + 2\sqrt{I_1/I_2}\cos\Delta\varphi \qquad (1-2)$$

When the optical path difference of coherent light is ΔL, there is another

$$\Delta\varphi = \frac{2\pi}{\lambda}\Delta L \qquad (1-3)$$

Then

$$I = I_1 + I_2 + 2\sqrt{I_1/I_2} \cos\left(\frac{2\pi}{\lambda}\Delta L\right) \qquad (1-4)$$

For a given ΔL, there is a cosine relationship between ΔL/λ and I, and the intensity and magnitude of the two beams of reflected light only affect the amplitude and contrast of the intensity of the reflected light I. In order to further illustrate the relationship between Δ L/λ and I, and then characterize the relationship between λ and I, the formula is simplified and assumed that $I_1 I_2$ I1 and I2 are approximately equal in size, i.e., I1=I2=I0, and the above equation can be simplified as:

$$I = 2I_0\left(1 + \cos\left(\frac{2\pi\Delta L}{\lambda}\right)\right) \qquad (1-5)$$

This equation shows that in ΔL/λ=0,0.5,1,1.5......, the extreme value of the reflected light intensity I can be obtained. When the initialization value of ΔL is in the μm order, the relationship between I and λ in the above equation is a cosine-like relationship, that is, the shape is much like a cosine, but as λ increases, the period of the signal broadens, as shown in the figure 1 C.

I1, I2 and π can be regarded as constants, and the coherent light intensity is only related to the optical path difference ΔL and the wavelength λ of the light.

It is easy to obtain the following equations (1-6) and (1-7).

$$\frac{\Delta L}{\lambda} = \pm k \qquad k=0,1,2,\ldots,\text{strengthen} \qquad (1-6)$$

$$\frac{\Delta L}{\lambda} = \frac{\pm(2k+1)}{2} \qquad k = 0,1,2,\ldots,\text{wane} \qquad (1-7)$$

If the white isocomplex color light is used and the spectrometer is used to receive the interference signal, the light intensity distribution map corresponding to the coherent light of different wavelengths can be obtained. With the change of wavelength, the interference signal will produce obvious intensity changes.

Then at the peak

$$\frac{\Delta L}{\lambda} = \pm k \qquad (1-8)$$

Considering the peak positions of two adjacent orders in the interference signal,

it can be obtained

$$\frac{2nd}{\lambda_1} = j \qquad (1-9)$$

$$\frac{2nd}{\lambda_2} = j+1 \qquad (1-10)$$

Among them, λ1 and λ2 are the wavelengths corresponding to the peaks of the two adjacent interferences (λ1>λ2), and j is the order.

Finishing is available

$$d = \frac{\lambda_1 \lambda_2}{2n(\lambda_1 - \lambda_2)} \qquad (1-11)$$

## 3. Instrument design

The construction of experimental instruments needs to meet:

(1) The output of the light source optical signal and the input of the feedback optical signal and the external optical signal;

(2) The light source is controllable;

(3) Real-time collection and processing of optical signals.

The main hardware is divided into three parts: a light source, a Y-type optical fiber with a probe, and a spectrometer.

The system construction uses the cold white light LED in the USB power supply port. In essence, the blue light LED is covered with phosphor, which has the characteristics of stable operation, low energy consumption, small volume and convenient use, As shown in Figure 2 A. The wavelength range is 420nm ~650nm, and the spectral curve of the light source is shown in the figure 2 B.

The purpose of optical fiber is the optical signal transmission unit. The optical signal transmission unit consists of a "Y" type optical fiber and a reflection probe, as shown in Figure 2C and Figure 2D. Considering the characteristics of each functional optical path of the system, the 6+1 optical fiber of the "Y" Raman system was used in the experiment, and the reflection probe matched with the optical fiber was used for the probe. The 6+1 fiber of the Y-Raman system has 7 fibers at the closing end, 6 fibers at one end and 1 fiber at the other end. The 6-core end is used to receive the light signal of the light source, and the 7-core end of the beam is equipped with a reflection probe to send out the light signal generated by the light source, and the light signal is irradiated on the test sample, and the light reflected by the surface is received by the probe, and then the reflected light signal is sent to the spectrometer by the 1 core end. The probe is a reflection probe used for standard measurements, which can effectively emit and receive optical signals.

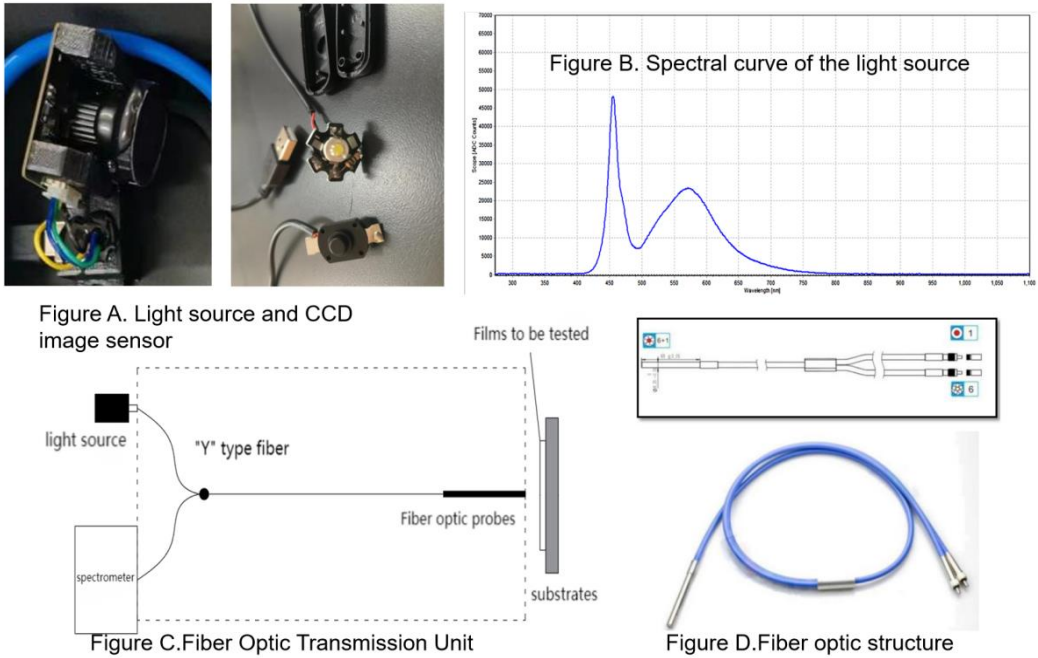

Figure 2

The optical signal processing work is mainly realized jointly by the slit grating, the CCD photoelectric image sensor and the computer processing system. The grating is used for dividing light and dividing the light source into spectra of different wavelengths onto the CCD detector; the CCD detector has many photosensitive units, each unit receives optical signals of a certain wavelength and transforms them into the corresponding electrical signal, and the computer data processing system is used to collect and analyze the electrical signals output by the CCD detector to obtain the spectral information of the sample. The spectrometer is designed with high light sensitivity and high resolution spectral information, which can detect very weak light signals and facilitate accurate analysis of the characteristics of the sample. The detection speed of CCD spectrometer is very fast, suitable for real-time analysis, bands can cover ultraviolet, visible, near-infrared and other bands, suitable for the analysis of a variety of samples.

The computational processing system is written in the VB programming language. The programming is based on open source data from the theremino website. The final design program system has important functions such as changing the integration time, sampling the spectral peak and wave trough, adjusting the imaging

Page 7

parameters, and exporting the spectral data,etc.

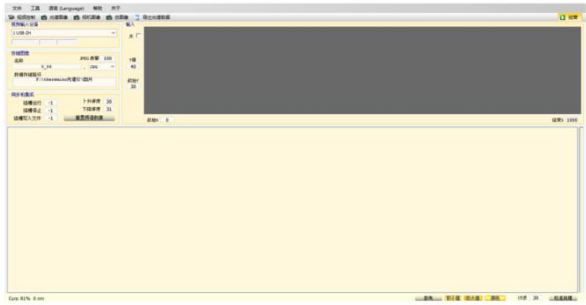 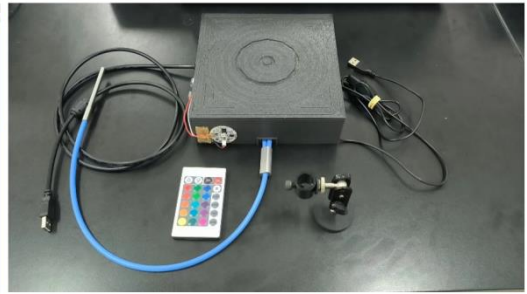

Figure A.Software pages　　　　　　　　Figure B.Instrument display

Figure 3

Figure 3 A and B shows the finished product display diagram and the software interface display diagram. The cover plate can be used as the experimental base for sample placement; the holder is used for fixing the optical fiber. The size parameters of the instrument shell are built based on the optical path. Reducing the scale of the realization of the optical path can make the instrument more portable. The overall 3D printing is a technology of making objects by layer based on digital model files.

## 4.Test verification

## 4.1 Data measurement

The film sample group is:

Group A: PI films with scalar thickness of 0.025mm and 0.05mm;

Group B: PI films with thickness scalars of 0.025mm, 0.05mm and 0.1mm;

Group C: PVC film with a thickness of approximately 0.025-0.100mm.

The spectra of interference light generated on each group of films are shown in Figure 4 A-F.

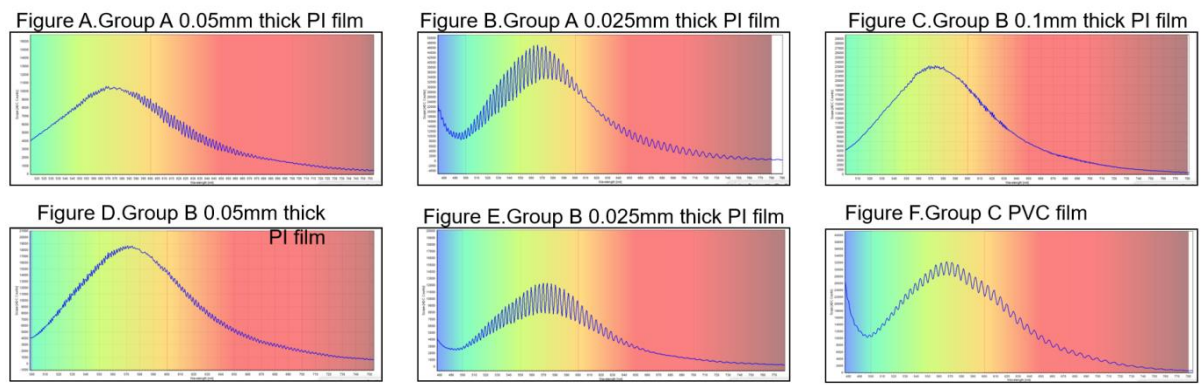

Figure 4

The spectral waveform diagram in the figure is sampled to find continuous and stable bands, and the two adjacent peaks (peaks) are extracted to calculate the film thickness. Then the sampling is repeated twice in other continuous and stable bands to calculate the average value, which is recorded in Table 1. Repeat the above procedure for sampling other films.

Table 1 Measurement data of film thickness

| The first time | | The second time | |
|---|---|---|---|
| sample | Calculated value （μm） | sample | Calculated value （μm） |
| 0.1PI film | 102.0 | 0.1PI film | 99.7 |
| 0.05PI film | 54.3 | 0.05PI film | 58.9 |
| 0.025PI film | 26.3 | 0.025PI film | 27.4 |
| 0.05PI film | 52.5 | 0.05PI film | 53.3 |
| 0.025PI film | 27.2 | 0.025PI film | 24.8 |
| PVC film | 19.3 | PVC film | 18.2 |
| The third time | | The fourth time | |
| sample | Calculated value （μm） | sample | Calculated value （μm） |
| 0.1PI film | 97.3 | 0.1PI film | 98.2 |
| 0.05PI film | 55.0 | 0.05PI film | 55.2 |
| 0.025PI film | 29.2 | 0.025PI film | 28.1 |
| 0.05PI film | 55.5 | 0.05PI film | 51.8 |
| 0.025PI film | 23.3 | 0.025PI film | 26.3 |
| PVC film | 17.6 | PVC film | 18.2 |
| The fifth time | | The sixth time | |
| sample | Calculated value （μm） | sample | Calculated value （μm） |
| 0.1PI film | 101.3 | 0.1PI film | 99.3 |
| 0.05PI film | 53.1 | 0.05PI film | 55.6 |
| 0.025PI film | 26.5 | 0.025PI film | 28.7 |
| 0.05PI film | 54.1 | 0.05PI film | 53.8 |
| 0.025PI film | 25.9 | 0.025PI film | 24.2 |
| PVC film | 19.5 | PVC film | 18.8 |

## 4.2 Data analysis

The data in Table 1 is used for calculation, and the film thickness can be calculated by taking into formula 1-10. The measurement error is systematically analyzed by using error percentage and relative uncertainty. The specific calculation and analysis process is as follows:

Calculating the mean film thickness and the arithmetic mean value of the measurement column can be obtained differently:

Table 2. Calculated data

| sample | 0.1PI film | 0.05PI film | 0.025PI film | 0.05PI film | 0.025PI film | PVC film |
|---|---|---|---|---|---|---|
| Calculated value (μm) | 99.6 | 55.4 | 27.7 | 53.3 | 25.3 | 18.6 |
| Standard deviation (μm) | 0.73 | 0.79 | 0.48 | 0.53 | 0.58 | 0.30 |

The synthetic standard uncertainty is

$$u_D = \sqrt{u_A{}^2 + u_B{}^2} = 0.3 \mu m$$

The relative uncertainty is calculated

$$E_D = \frac{u_D}{\bar{D}} \times 100\% = 0.3\%$$

The extended uncertainty of the measurement result with confidence P = 0.95 is

$$U_{0.95} = \sqrt{(t_P u_A)^2 + u_B{}^2} = 0.77 \mu m$$

It can be seen that the measurement has higher accuracy and less interference by external factors.

Result analysis: Due to the influence of noise and the characteristics of different bands, different wave crest sampling points have different influences on the calculation, so the sampling and scalar difference of some wave peaks is large, but the overall maximum error is not more than 15%. The final data measured by the experiment is very close to the scalar as a whole, with an error of less than 4 microns.

The average total error is 5%, which has a good precision. In addition, the measurement accuracy of 0.1mm thick film is the highest, the error is maintained at 1%, the thinner the film measurement, on the one hand, by external factors, on the one hand, the experiment itself is limited in measuring capacity, so the measurement accuracy is lower, these are completely improved under the conditions of human intervention.

# 5.Conclusion

Result analysis: Due to the influence of noise and the characteristics of different bands, different wave crest sampling points have different influences on the calculation, so the sampling and scalar difference of some wave peaks is large, but the overall maximum error is not more than 15%. The final data measured by the experiment is very close to the scalar as a whole, with an error of less than 4 microns. The average total error is 5%, which has a good precision. In addition, the measurement accuracy of 0.1mm thick film is the highest, the error is maintained at 1%, the thinner the film measurement, on the one hand, by external factors, on the one hand, the experiment itself is limited in measuring capacity, so the measurement accuracy is lower, these are completely improved under the conditions of human intervention.

For experimental validation, we perform membrane thickness measurements on PI and PVC films of different thicknesses, calculate the film thickness from the collected interference signals and compare it with the theoretical values. The experimental results show that the measurements of the membrane thickness instrument are highly consistent with the theoretical value, and the total error percentage and relative uncertainty are within the acceptable range. Especially for the PI film, whether the sample is 0.1 μm, 0.05 μm or 0.025 μm, the instrument can measure its thickness accurately, and the measurement results are stable and reliable. Moreover, the instrument also has good reproducibility and reproducibility with small deviation between multiple measurements, further demonstrating the accuracy and reliability of its measurements.

Besides the measurement accuracy, the non-contact membrane thickness instrument also has the significant advantages of convenient operation, sample protection and fast measurement speed. Because the instrument adopts non-contact measurement method, it avoids the risk of damage and contamination to the sample, which is especially suitable for the sample surface quality. At the same time, the measurement speed of the instrument is fast, which can quickly complete the measurement work of multiple samples, greatly improving the measurement efficiency. In addition, the instrument shell is small and portable, the whole appears black to prevent the influence of external natural light on the internal instrument, and the material is environmental protection, excellent performance, high strength, high toughness, impact resistance and heat resistance, further enhance the practicability and durability of the instrument.

In conclusion, the contactless membrane thickness instrument designed in this study has wide application prospects and potential economic value in the field of membrane thickness measurement. In the future, we will continue to optimize the instrument design to improve its measurement accuracy and stability, while exploring more application scenarios to meet the needs of different industries for membrane thickness measurement.

**Acknowledgments.** National Natural Science Foundation of China (62305262). Shaanxi Fundamental Science Research Project for Mathematics and Physics (22JSQ026). Shaanxi Province College Student Innovation and Entrepreneurship Training Program Project (202410703052).

**Competing interests.** The authors declare that they have no competing interests.

**Data availability.** Data underlying the results presented in this paper are not publicly available at this time but may be obtained from the authors upon reasonable request.